\documentclass[letterpaper, 10 pt,romanappendices, conference]{ieeeconf}
\IEEEoverridecommandlockouts  
\usepackage{textcomp}

\usepackage{amsmath,amssymb}

\usepackage{amsthm}
\usepackage{amsfonts,mathrsfs}
\usepackage{color}
\usepackage{graphicx}
\usepackage{epsfig}
\usepackage{subcaption}

\usepackage{algorithm,algorithmicx}
\usepackage{algpseudocode}
\usepackage[hidelinks]{hyperref}
\usepackage{marginnote}
\usepackage{empheq}
\usepackage{bm}
\usepackage{accents}
\usepackage{cite}
\usepackage{balance}
\usepackage{latexsym}
\usepackage{graphicx}
\usepackage{float}
\usepackage{colortbl}
\usepackage{hyperref}
\let\labelindent\relax\usepackage{enumitem}
\usepackage{multicol}
\usepackage{float}
\usepackage{cite}
\usepackage{multirow}
\usepackage[normalem]{ulem} 

\usepackage[usenames,dvipsnames,svgnames,table,xcdraw]{xcolor}

\usepackage{setspace}

\newcommand{\citep}{\cite}

\DeclareMathOperator{\proj}{proj}

\usepackage{soul}

\newcommand{\bs}{\boldsymbol}
\newcommand{\mc}{\mathcal}
\newcommand{\bb}{\mathbb}
\newcommand{\R}{\bb R}
\newcommand{\N}{\mathbb{N}}

\let\L\relax\newcommand{\L}{\mathcal{L}}

\DeclareMathAlphabet{\mathbbmsl}{U}{bbm}{m}{sl}

\newcommand{\dom}{\operatorname{dom}}

\newcommand{\argmin}{\operatorname{argmin}}

\newcommand{\fix}{\operatorname{fix}}

\newcommand{\Id}{\mathrm{Id}}

\newcommand{\diag}{\operatorname{diag}}

\newcommand{\col}{\operatorname{col}}
\newcommand{\zer}{\operatorname{zer}}

\newcommand{\nc}{\mathrm{N}}
\newcommand{\0}{\mathbf{0}}
\newcommand{\1}{\mathbf{1}}

\newcommand{\gph}{\operatorname{gph}}

\newcommand{\Rmnum}[1]{\expandafter\@slowromancap\romannumeral #1@}

\newcommand{\qedd}{\ensuremath{\hfill \blacksquare}}

\newcommand{\bx}{\boldsymbol{x}}
\newcommand{\by}{\boldsymbol{y}}

\newcommand{\blambda}{\boldsymbol \lambda}
\newcommand{\bnu}{\boldsymbol \nu}

\newcommand{\SOL}{\mathrm{SOL}}

\newcommand{\bOmega}{\boldsymbol{\Omega}}

\newcommand{\bomega}{\bm{\omega}}

\newcommand{\X}{\bs{\mc X}}

\newcommand{\bz}{\boldsymbol{z}}

\usepackage{changepage}

\makeatletter
\newcommand{\specialcell}[1]{\ifmeasuring@#1\else\omit$\displaystyle#1$\ignorespaces\fi}
\makeatother
\newtheorem{assumption}{Assumption}
\newtheorem{proposition}{Proposition}

\newtheorem{definition}{Definition}
\newtheorem{lemma}{Lemma}

{\it}{}

\newtheorem{remark}{Remark}

\newcommand{\VI}{\mathrm{VI}}

\title{ A {Semi-Decentralized} Tikhonov-based Algorithm for {Optimal} Generalized Nash Equilibrium Selection }
\author{Emilio Benenati, Wicak Ananduta, and Sergio Grammatico 
\thanks{The authors are with the Delft Center for Systems and Control (DCSC), TU Delft, the Netherlands. E-mail addresses: \texttt{\{e.benenati, w.ananduta, s.grammatico\}@tudelft.nl}. }
\thanks{This work was partially supported by the ERC under research project COSMOS (802348). }
}

\allowdisplaybreaks

\begin{document}
	
	\maketitle
	\begin{abstract} To optimally select a generalized Nash equilibrium, in this paper, we propose a semi-decentralized algorithm based on a double-layer Tikhonov regularization method. Technically, we extend the Tikhonov method for equilibrium selection in non-generalized games {to the generalized case} by {coupling it with} the preconditioned forward-backward splitting, which guarantees linear convergence to the solutions of the inner layer problem {and allows for a {semi}-decentralized implementation.} We then {establish} a conceptual connection and draw {a} comparison between the proposed algorithm and the hybrid steepest descent method, the other {known} distributed framework {for solving the selection} problem. 
	\end{abstract}

	\section{Introduction}
	{Several} multi-agent decision processes can be modelled as a game, that is, a set of inter-dependent optimization problems. In particular, if the agents are coupled not only through their respective objective functions, but also through a shared constraint set, then we label the setting {as} a \emph{generalized} game. Application examples  for generalized games include traffic routing \cite{bakhshayesh_decentralized_2021}, peer-to-peer energy markets \cite{belgioioso_operationally-safe_2022,ananduta_equilibrium_2022} and cognitive radio networks \cite{wang_generalized_2014}. A typical solution paradigm is the generalized Nash equilibrium (GNE), that is, an optimal situation for each agent given the decisions of the remaining agents, and especially the sub-class of variational GNEs (v-GNES), which has recently received widespread attention {due to its stability properties \cite{facchinei_generalized_2010}.} 
	
	{Plenty} efficient v-GNE seeking algorithms, e.g. \cite{yin_nash_2011,kannan_distributed_2012,grammatico_proximal_2018,yi_operator_2019, franci_distributed_2020,belgioioso_distributed_2020}, have been developed  for games {that} satisfy a monotonicity condition. Crucially, {monotone} games admit in general an infinite number of v-GNEs (unless a much more restrictive \emph{strong} monotonicity condition is imposed). {An appealing method to deal with the non-uniqueness of the solution is to select {a} GNE {that} optimizes some desirable {objective,} as proposed in \cite{scutari_equilibrium_2012,facchinei_vi-constrained_2014,benenati_optimal_2022,benenati_optimal_2022-1}. 
	{The selection algorithms in} \cite{scutari_equilibrium_2012,facchinei_vi-constrained_2014}  {use Tikhonov's regularization method} and builds on the literature of Variational Inequalities (VIs) to cast the selection problem as a VI-constrained VI, which is solved by finding {a sequence of approximate solutions to regularized games.} The algorithms in our previous works \cite{benenati_optimal_2022, benenati_optimal_2022-1}, instead, rely on fixed point selection theory  and {use the {hybrid steepest descent method} (HSDM)  \cite{yamada_hybrid_2005}, which pairs}  an appropriate nonexpansive operator {with a gradient descent.} While the latter is recently proposed as distributed algorithms for GNE selection, the former {works for} non-generalized games {only.}
	
	In this paper, we {close} the gap between the two solution frameworks by devising a Tikhonov-based algorithm  for equilibrium selection in generalized, monotone games. {Similarly to {\cite{facchinei_vi-constrained_2014},} we cast the GNE selection problem as a VI-constrained VI.} Then, we design a decentralized algorithm based on the preconditioned Forward Backward (pFB) \cite{yi_operator_2019} to solve each regularized sub-problem with linear convergence rate. The proposed algorithm requires 
	a central coordinator 
	to compute the gradient of the selection function and to check the termination condition for the pFB algorithm. {Compared with the Tikhonov-based GNE seeking algorithms in \cite{yin_nash_2011,kannan_distributed_2012} {that compute} the minimum-norm v-GNE, our {proposed} algorithm works {for} general convex selection functions.} 
	  Interestingly, we {also} find a theoretical connection between the proposed Tikhonov method and the HSDM. Although neither method generalizes the other, {the HSDM} can be cast as a forward-backward step towards the solution of the Tikhonov regularized {inner layer} problem {with an appropriate choice of the nonexpansive operator.}  {In Section \ref{sec:num_sim},} we {present} a numerical comparison between the two methods. 
	\vspace{-1pt}
\paragraph*{Notation} The Euclidean inner product and norm are denoted  respectively by $\langle x, y \rangle$  and $\|\cdot\|$. For a {symmetric} matrix $\Psi\succ 0$, we denote the $\Psi$-induced norm {by} $\|\cdot\|_\Psi$ and {define} $\langle x, y \rangle_\Psi = \langle x, \Psi y\rangle$. Nonlinear set-valued operators are denoted in calligraphic letters, e.g. $\mc T:\R^n \rightrightarrows \R^n$. For a matrix $A$, $\|A\|$ denotes its spectral norm.  {We denote the vector of all $1$ ($0$) with dimension $n$ by $\1_n$ ($\bm 0_n$). } The column stack operation is denoted {by} $\col(\cdot)$. 
\vspace{-3pt}
\paragraph*{Operator theory}
We denote by $\Id$ the identity operator. We denote the normal cone of a closed convex set $C$ as $\nc_C(\cdot):\R^n\rightrightarrows\R^n$ \cite[Def. 6.38]{bauschke_convex_2017} and $\proj_C(x)=\argmin_{z\in C} \| x-z \|$. {For an operator $\mc T \colon \bb R^n \rightrightarrows \bb R^n$, we denote its zero set by $\zer(\mc T) := \{x \in \dom(\mc T) \mid 0 \in \mc T(x)\}$, its fixed point set by $\fix(\mc T) := \{x \in \dom(\mc T) \mid  x\in \mc T(x) \}$ {and its resolvant  {by} {$\mc J_{\mc T}:=(\Id +  \mc T)^{-1}$}} \cite[Def. 23.1]{bauschke_convex_2017}.} We refer to an operator $\mc T:C\rightrightarrows\R^n$ as: \\
\emph{($\mu$-strongly) monotone}:  if,  for any {two pairs  $(x,y), (x',y')\in \gph(\mc T):= \{ (x,y)| x \in C, y\in \mc T(x)\},$ it holds that} $\langle y-y', x-x'\rangle \geq \mu\| x - x'\|^2$, with $\mu\geq 0$ ~($>0$); \\ 
{\emph{Monotone-plus}:  if it is monotone and, for any two pairs $(x,y), (x',y')\in \gph(\mc T)$ such that  $\langle y-y', x-x'\rangle =0$, it holds $(x,y')\in\gph(\mc T)$ and $(x',y)\in\gph(\mc T)$;} \\ 
\emph{Lipschitz continuous}:  if there exists $L>0$, such that, for all $x, x' \in C$, 
$ \| \mc T(x) - \mc T(x')\| \leq L \|x-x'\|$; \\
\emph{Nonexpansive}:  if Lipschitz continuous with $L=1$; \\
\emph{Contractive}:  if Lipschitz continuous with $L<1$; \\
\emph{$\alpha$-averaged nonexpansive}: if  for $\alpha \in (0,1)$ there exists $\mc R \colon C \to \bb R^n$ nonexpansive such that $\mc T = (1-\alpha)\Id + \alpha \mc R$.
\vspace{-3pt}
\paragraph*{Variational inequalities (VIs)} For a closed convex set $K$ and $\mc T: K\to K$, we denote by $\VI(K, \mc T)$ the problem 
\begin{equation*}
	\text{find}~\bx^*\in K ~\text{s.t.} ~ \inf_{\bx \in K}\langle \mc T(\bx^*), \bx - \bx^*\rangle \geq 0,
\end{equation*}
{and its} solution set {by}
\begin{equation*}
	\SOL(K,\mc T):=\{\bx^*\in K | \bx^* ~\text{solves}~ \VI(K, \mc T)\}.
\end{equation*}%
{A point $\hat{\bx}$ is an $\varepsilon$-approximate solution of $\VI(K,\mc T)$ if $$\varepsilon \geq \textstyle \inf_{\bx \in \SOL(K,\mc T)} \| \bx - \hat{\bx}\|,$$ where $\varepsilon\geq 0$ denotes the approximation error. }

\section{Equilibrium selection as a {variational inequality problem}}
\label{sec:eq_select}
	We consider the multi-agent decision process in which each of $N$ agents aims at solving an optimization problem over the decision variables $x_i\in\mc X_i \subset \R^{n_i}$, where $i\in\{1,..., N\}=:\mc I$. Let us denote $n:=\sum_{i\in\mc I} n_i$ and $\mc I_{-i}:=\mc I \setminus \{i\}$. Crucially, the decision problem associated with agent $i$ is coupled to the decision variables of the remaining agents $\bx_{-i}:=(x_j)_{j\in \mc I_{-i} }$ both through the objective function $J_i:\R^n\to\R$, {where $n = \sum_{{i}\in \mc I} n_i$, and some}  constraints. We consider the case in which the coupling constraints  are defined via  $m\in\N$ inequalities of the form
	\begin{equation}
		\label{eq:coupling_constr}
		\textstyle\sum_{{i}\in \mc I} A_i x_i  \leq b,
	\end{equation}
	where {$A_i \in \R^{m\times n_i}$}, for each $i\in\mc I$, { and $b \in \R^m$.} This problem is commonly referred to as a \emph{generalized game} and we formalize it as follows:
		\begin{subequations}
		\begin{empheq}[left={\forall i \in \mc I \colon \empheqlbrace\,}]{align}
			\underset{x_i \in \mc X_i}{\operatorname{min}} \quad & J_i(x_i, \bm x_{-i}) \label{eq:game:cost}\\
			\operatorname{s.t.} \quad & A_i x_i \leq b - \textstyle\sum_{{j}\in\mc I_{-i} } A_j x_j. \label{eq:game:constr}
		\end{empheq}
		\label{eq:game}%
	\end{subequations}%
{We note that we can reformulate the more general case with separable convex coupling constraint functions into the form of \eqref{eq:game} by augmenting the decision variables \cite[Rem. 2]{grammatico_proximal_2018}.} Let us now denote $\X := \Pi_{i\in\mc I} \mc X_i$ and the collective feasible set of the game in \eqref{eq:game} as
\begin{equation*}
	\Gamma:= {\X} \cap \{\bx~|~ \eqref{eq:coupling_constr} ~\text{holds true}\}. 
\end{equation*} 
	As a solution paradigm for the problem in \eqref{eq:game}, we {consider} the set of generalized Nash equilibria, that is, a set of points from which no agent has any incentive to unilaterally deviate:
	
	\begin{definition}
		A set of decision variables $\bx^*\in\X$ is a GNE for the game in \eqref{eq:game} if, for each $i\in\mc I$,
		\begin{equation*}
			J_i(x_i^*, \bx^*_{-i}) \leq J_i(x_i, \bx^*_{-i}),
		\end{equation*}
		for any $x_i\in\mc X_i \cap \{{y \in \R^{n_i}}~|~ A_i {y} \leq -\sum_{j\in{\mc I_{-i}}}A_jx_j^{{*}}\}$.
	\end{definition}
	The following assumptions, {which  are standard for  GNE seeking problems \cite{franci_distributed_2020,belgioioso_distributed_2020}, guarantee the existence of a GNE \cite[Prop. 12.11]{palomar_convex_2009}:
	\begin{assumption}
		\label{as:local_objective}
		For each $i$, $J_i(\cdot, \bx_{-i})$ is convex and continuously differentiable for any $\bx_{-i}$.
	\end{assumption}
\begin{assumption}
	\label{as:constraints}
	For all $i\in\mc I$, $\mc X_i$ is compact and convex; $\Gamma$ is non-empty and it satisfies Slater's constraint qualification.
\end{assumption}
\begin{assumption}
	\label{as:monotonicity}
	The pseudogradient mapping of the game in \eqref{eq:game}, defined as
	\begin{equation*}
		F(\bx):=\col((\nabla_{x_i} J_i(x_i, \bx_{-i}))_{i\in\mc I}),
	\end{equation*}
is monotone and $L_F$-Lipschitz continuous.
\end{assumption}
{For solving} the problem in \eqref{eq:game} via a {decentralized} algorithm, {we assume that the agents can {exchange information} over an undirected, connected {communication network. We denote} the set of neighbours of agent $i$ {in this network} by  $\mc N_i$.} {For simplicity,} {we consider the case where, for all $i\in\mc I$, $J_i$ depends on {$x_i$ and the decision variables of (a subset of) $\mc N_i$,} so that each agent is able to evaluate its cost function by receiving the decision variables implemented by some neighbouring agents.} {Furthermore,} each agent maintains a local estimate of the dual variable associated {with} the shared constraints in \eqref{eq:coupling_constr}, {denoted by $\lambda_i$,} as well as an auxiliary variable $\nu_i$, which is communicated to the other agents in order to achieve consensus over the dual variable {estimates.}  We then define the following extended {Karush-Kuhn-Tucker} (KKT) operator, first proposed in \cite{yi_operator_2019}, which includes both the optimality conditions for the coupled optimization problems in \eqref{eq:game} and the consensus condition:
\begin{align}
	\begin{split}
		\label{eq:KKT_op}
			\mc T^{\text{KKT}}(\bomega) =& \mc A(\bomega) + \mc B(\bomega) + \mc C(\bomega),\\
			\mc A(\bomega) :=& {\nc_{\X}(\bx) \times
			\nc_{\R^{|\mc I|m}_{\geq 0}}(\blambda) \times 
			\{\bs 0_{|\mc I|m}\} } , \\
			\mc B  :=&  \begin{bmatrix}
			{F}(\bx) \\
			- \bar\L\blambda \\
			0
		\end{bmatrix}, \  \mc C(\bomega):= \begin{bmatrix}\bs A^{\top}\blambda  \\   b-\bs A \bs x  - \bar\L\bnu  \\  \bar\L\blambda \end{bmatrix},
	\end{split}
\end{align}
where {$\bomega =\col(\bx, \blambda, \bnu)$ and} $\bar\L = \L\otimes I_m$ and $\L$ is the Laplacian matrix of the communication graph. It is well known that a subset of the GNEs of \eqref{eq:game} can be found as the zero set  of {$\mc T^{\text{KKT}}$} \cite[Thm. 2]{yi_operator_2019}. {Such GNEs are known as} the variational GNEs (v-GNEs) of the game in \eqref{eq:game} \cite[Def. 3.10]{facchinei_generalized_2010}. {This class of solutions} has been widely studied and multiple efficient algorithms for their computation exist. However, {since} Assumptions \ref{as:local_objective}--\ref{as:monotonicity} are not enough to guarantee that $\zer(\mc T^{\text{KKT}})$ is a singleton, most of the algorithms in the literature resort to computing an unspecified v-GNE among the possibly infinitely many.

Instead, we develop an algorithm to compute an optimally selected v-GNE according to the selection function $\phi$, in the sense that we aim at solving the selection problem
\begin{align}
	\label{eq:GNE_selection_problem}
	\min_{\bomega} \phi(\bomega) \quad \operatorname{s.t.} \ \bomega\in\zer(\mc T^{\text{KKT}}).
\end{align}
{Let us} {postulate} the following assumptions on the selection function:
\begin{assumption}
	\label{as:phi}
	{The selection function} $\phi$ is convex, continuously differentiable, and coercive. {Furthermore, its gradient} is $L_{\nabla \phi}$-Lipschitz continuous.
\end{assumption}
By \cite[Prop. 23.39]{bauschke_convex_2017}, $\zer(\mc T^{\text{KKT}})$ is a closed convex set under Assumptions \ref{as:local_objective} and \ref{as:monotonicity}. Therefore, \eqref{eq:GNE_selection_problem} is a convex optimization problem. However,  $\zer(\mc T^{\text{KKT}})$ can seldom be written in {a} closed form and, thus, \eqref{eq:GNE_selection_problem} cannot be solved by standard optimization algorithms. To derive an algorithmic solution, we further massage its formulation by noting that $\zer(\mc T^{\text{KKT}})$ is the solution set to a Variational Inequality (VI) \cite[Sec. 1.3.1]{facchinei_finite-dimensional_2003}. Specifically, by defining $\bOmega:= \X \times \R_{\geq 0}^{Nm} \times \R^{Nm} $ and by {\cite[Eq. (1.1.3)]{facchinei_finite-dimensional_2003}},
\begin{equation}\label{eq:kkt_solves_vi}
	\bomega^*\in\zer(\mc T^{\text{KKT}}) \Leftrightarrow \bomega^*\in\SOL(\bOmega, \mc B + \mc C).
\end{equation}
{Following the equivalence between convex optimization problems and VIs \cite[Sec. 1.3.1]{facchinei_finite-dimensional_2003}, we then} recast \eqref{eq:GNE_selection_problem} as the {following} VI-constrained VI:
\begin{equation}
	\label{eq:vi_constr_vi}
	{\VI(\SOL(\bOmega, \mc B + \mc C), \nabla\phi)}.
\end{equation}
{As discussed in \cite{facchinei_vi-constrained_2014},} {we can solve} {\eqref{eq:vi_constr_vi}} by finding the {$\varepsilon_k$-approximate} solutions  to a sequence of regularized sub-problems, {indexed by $k\in\N$}: 
	\begin{align}
			&\VI(\bOmega, \mc{B} + \mc C + \gamma_k \nabla\phi + \alpha(\Id - \bomega)),  ~~\bomega\in\bOmega, \label{eq:regularized_VI}
	\end{align}
where the sequence of regularization weights $\alpha$, $(\gamma_k)_{k\in\N}$ and approximation errors $(\varepsilon_k)_{k\in\N}$ are chosen according to the following criteria:

\begin{assumption} \label{as:stepsize_regul} {The parameter $\alpha$ is positive;} $(\gamma_k)_{k\in\N}$ and  $(\varepsilon_k)_{k\in\N}$ are positive and non-negative sequences of real numbers, respectively, such that $ \textstyle\sum_{k\in\N} \gamma_k = \infty $ and eventually $\varepsilon_k = 0.$
\end{assumption} 

{{By denoting} ${\bar\gamma}:=\sup_{k\in\N}\gamma_k$,  we {observe} the following properties of the operators that define \eqref{eq:regularized_VI}:
 \begin{lemma}\label{le:str_monotone}
 	Let Assumption{s} \ref{as:monotonicity}--\ref{as:stepsize_regul} hold true. For any {$k\in\N$}, $\alpha>0$, {and} $\bomega\in\bOmega$: 
 	\begin{enumerate}
 		\item $\mc{B} + \mc C + {\gamma_k} \nabla\phi + \alpha(\Id - \bomega)$ is $\alpha$-strongly monotone,
 		\item  $\mc{B}+ {\gamma_k} \nabla\phi + \alpha(\Id - \bomega)$ is $L_G$- Lipschitz continuous, where $L_G:=\max(L_F, 2|\mc N_i|_{i\in\mc I}) + {{\bar\gamma}}L_{\nabla\phi} + \alpha$.
 	\end{enumerate}
 \end{lemma}
 
  By applying \cite[Cor. 2.2.5]{facchinei_finite-dimensional_2003} and \cite[Thm. 2.3.3a]{facchinei_finite-dimensional_2003}, we conclude that the regularized problem in \eqref{eq:regularized_VI} admits one and only one solution. The following result, which follows directly from \cite[Thm. 2]{facchinei_vi-constrained_2014}, formalizes a prototypical algorithmic solution to the problem in \eqref{eq:vi_constr_vi} based on a sequential solution of the problem in \eqref{eq:regularized_VI}.
\begin{proposition} \label{le:Tikhonov_alg}
	Let Assumptions \ref{as:local_objective}--\ref{as:stepsize_regul} hold. {Let $(\bomega^{(k)})_{k \in \N} \in \bOmega$ and, for every $k$, $\bomega^{(k+1)}$ be the $\varepsilon_k$-approximate solution of the VI in (7) with $\bomega=\bomega^{(k)}$. } 
		Then, the sequence $(\bomega^{(k)})_{k\in\N}$	is bounded and each of its limit points is a solution of {\eqref{eq:vi_constr_vi}}.
		
\end{proposition}

\section{Distributed equilibrium selection}
{In view of  Proposition \ref{le:Tikhonov_alg}, {next} we derive a {semi-decentralized} algorithm for recursively generating a sequence $(\bomega^{(k)})_{k\in\N}$ such that, for all $k$, $\bomega^{(k+1)}$ is a $\varepsilon_k$-approximate solution to \eqref{eq:regularized_VI} with $\bomega=\bomega^{(k)}$. We note that the exact solution, {denoted by} $\bomega^*_k$,  equivalently satisfies the monotone inclusion}
\begin{align}
	\label{eq:regul_monotone_incl}
	\begin{split}
&	0 \in (\mc A + \mc B + \mc C +\gamma_k \nabla\phi + \alpha(\Id - \bomega^{(k)}))(\bomega_k^*).
\end{split}
\end{align}
We then {apply the preconditioned forward-backward (pFB) method on \eqref{eq:regul_monotone_incl},} proposed in \cite{yi_operator_2019} in the context of decentralized GNE seeking, for {computing $\bs\omega^*_k$.} {To this end,} let us define the following matrix, which collects the step sizes of the proposed algorithm:
\begin{equation}
	\Psi= 
		\diag(\bs\rho^{-1},  \bs\tau^{-1}, \bs\sigma^{-1}),
\end{equation}
where $\bs\rho :=\diag(\rho_i I_{n_i})_{i\in\mc I}$, $\bs\tau :=\diag(\tau_i I_{n_i})_{i\in\mc I}$, $\bs\sigma :=\diag(\sigma_i I_{n_i})_{i\in\mc I}$ collect the step sizes {associated with} the primal, dual and auxiliary variables, respectively. {Let us} make the following {design choice} on the matrix $\Psi$:
\begin{assumption} 
	\label{as:stepsizepFB}
	Let 
	\begin{align*}
	& r^\mathrm{x}_i:= \textstyle\max_{j=1,\dots,n_i} \sum_{k=1}^m |[A_i]_{jk}|,  \\
	& r^\lambda_i:=\textstyle \max_{j=1,\dots,n_i} \sum_{k=1}^m |[A_i]_{jk}| + 2|\mc N_i^\lambda|, \\
	& r^\nu_i := 2|\mc N_i^\lambda|.
	\end{align*}
	{Furthermore,} let $r = \max_{i\in\mc I}(r^\mathrm{x}_i, r_i^\lambda, r_i^\nu)$ {and} $\delta > \max(\frac{L_G^2}{\alpha}, 2r )$, {with $L_G$ defined as in Lemma \ref{le:str_monotone}.} For all $i \in \mc I$:
	\begin{enumerate}
		\item $( 2\delta - r^\mathrm{x}_i)^{-1}\leq \rho_i \leq ( \delta + r^\mathrm{x}_i)^{-1}$;
		\item $(2\delta - r^\lambda_i)^{-1} \leq \tau_i  \leq (\delta + r^\lambda_i)^{-1}$;
		\item  $(2\delta  - r^\nu_i)^{-1}\leq \sigma_i \leq (\delta  +r^\nu_i)^{-1}$.
	\end{enumerate}

\end{assumption}
{Let us} also define the preconditioning matrix
\begin{equation}
	\Phi =  \Psi+
	\begin{bmatrix}
		\bs 0 & -\bs A^{\top} & \bs 0 \\
		-\bs A & \bs 0 & - \bar\L  \\
		\bs 0 & -  \bar\L & \bs 0
	\end{bmatrix}.
\end{equation} 
\begin{lemma}
	\label{le:psi_matrix}
	Under Assumption \ref{as:stepsizepFB}, $\Phi\succcurlyeq \delta I$ and $\frac{\delta}{\|\Phi\|}\geq\frac{1}{2}$.
\end{lemma}

The pFB operator for the inclusion in \eqref{eq:regul_monotone_incl} reads, for all $k\in\N$, as
\begin{align}
	\label{eq:pFB}
	\begin{split}
			\mc T^{\text{pFB}}_{{k}} = &(\Id + \Phi^{-1} (\mc A + \mc C))^{-1} \\
			& (\Id - \Phi^{-1}(\mc B + \gamma_k\nabla\phi +\alpha(\Id - \bomega^{(k)}) ) )).
	\end{split}
\end{align}
The following result formalizes the convergence of the fixed-point iteration generated by $\mc T^{\text{pFB}}_{k} $ to the solution of \eqref{eq:regularized_VI}.
\begin{lemma}
	\label{le:inner_loop_convergence}
	Let Assumptions \ref{as:local_objective}--\ref{as:stepsizepFB} hold true. Let $\by^{0}\in\bOmega$. {Then, for all $k\in\N$,} the sequence $(\by^{(t)})_{t\in\N}$ generated by the {fixed-point} iteration 
	\begin{equation}\label{eq:pFB_iteration}
		\by^{(t+1)} = \mc T^{\mathrm{pFB}}_{k} (\by^{(t)}),
	\end{equation}
	{where $\mc T^{\text{pFB}}_{{k}}$ is defined in \eqref{eq:pFB},} converges linearly in the $\Phi$-induced norm to {$\bomega^*_k$} {in \eqref{eq:regul_monotone_incl}} and
	\begin{equation}\label{eq:error_bound}
		\|\by^{(t)} - \bomega^*_k \|_{\Phi}  \leq \|\by^{(t+1)} - \by^{(t)}\|_{\Phi} /(1-\beta),
	\end{equation}
with $\beta = (1 + \tfrac{L_G^2}{\delta^2} - \tfrac{2\alpha}{\|\Phi\|}) $.
\end{lemma}

\begin{remark}
	{Differently from the pFB operator in \cite{yi_operator_2019}, that in \eqref{eq:pFB} has additional regularization terms, involving the gradient of the selection function, and {thus achieves the desired} linear rate.}
\end{remark}
\begin{algorithm}[t]
	\caption{Tikhonov-regularized-pFB for optimal GNE  selection}
	\label{alg:FBF_tich}
	\textbf{Initialization.} Let $\alpha$, $(\varepsilon_k)_{k\in\N}$ and $(\gamma_k)_{k\in\N}$ satisfy Assumption \ref{as:stepsize_regul}. Let $\bomega^{(0)} \in \Pi_{i\in\mc I} ({\mc X_i})  \times  {\R_{\geq 0}}^{|\mc I|m} \times \R^{|\mc I| m}$. Let  $\bs\rho, \bs\sigma, \bs\tau$ satisfy Assumption \ref{as:stepsizepFB}.
	\\
	\textbf{Outer iteration: for $k\in\N_0$}
	\begin{enumerate}[leftmargin=*]
		\item Each agent $i\in\mc I$ sets:
		\begin{equation}
			 \hspace{-2pt} ({x}^{(k,0)}_i, {\lambda}^{(k,0)}_i, {\nu}^{(k,0)}_i)  \hspace{-2pt} \leftarrow  \hspace{-2pt} y_i^{(k,0)}  \hspace{-2pt} \leftarrow  \hspace{-2pt} \omega_i^{(k)}.
		\end{equation}
		\item \textbf{Inner iteration:  for $t\in\N_0$}\\
		\textbf{-For each agent $i \in \mc I$:}
		\begin{enumerate}[leftmargin=*]
			\item Receive $x_j^{(k,t)}$, $\lambda_j^{(k,t)}, \nu_j^{(k,t)}$ from agent $j \in \mc N_i$ and $\nabla\phi_{\omega_i}(\by^{(k,t)}) $ from the aggregator.
			\item Update:
			\begin{align}
				\begin{split}				\label{eq:primal_update}
					&\hspace{-20pt}	x_i^{(k,t+1)} =\hspace{-1pt}  {\proj_{\mc X_i}}\hspace{-0pt}\left[{x}_i^{(k,t)} \hspace{-1pt}- \hspace{-2pt} \rho_i \big(\nabla_{x_i}{J}_i(\bx^{(k,t)}) \hspace{-0pt} + \right. \\
					&\left. \hspace{-20pt} ~\hspace{-2pt}A_i^\top \hspace{-1pt} \lambda_i^{(k,t)} + \gamma_k \nabla_{x_i} \phi(\by^{(k,t)}) + \alpha(x_i^{(k,t)} - {x}^{(k,0)}_i )\right]\hspace{-2pt},
				\end{split}
			\end{align}
		\begin{align}
				\begin{split}				\label{eq:aux_update}
				\hspace{-20pt} \nu_i^{(k,t+1)} &= \nu_i^{(k,t)} - \sigma_i\left( \textstyle\sum_{j \in \mc N_i} (\lambda_i^{(k,t)}- \lambda_j^{(k,t)} ) \right. \\
			  & \left. +\alpha(\nu_i^{(k,t)} - {\nu}^{(k,0)}_i ) + \gamma_k \nabla_{\nu_i}\phi(\by^{(k,t)}) \right).
				\end{split}
			\end{align}
			\item Receive $ \nu_j^{(k,t+1)}$ from agent $j \in \mc N_i$.
			\item Update:
			{\small
			\begin{align}
				\label{eq:dual_update}
					\begin{split}
					&\hspace{-30pt}\lambda_i^{(k,t+1)} = \proj_{\R^m_{\geq 0}}\Big[\Big.\lambda_i^{(k,t)} + \tau_i\Big(\Big. \alpha(\lambda_i^{(k,t)} - {\lambda}^{(k,0)}_i )- b_i  \\
					&\hspace{-30pt}+ A_i(2 x_i^{(k,t+1)} - x_i^{(k,t)})  + \Big.\Big.\textstyle\sum_{j \in \mc N_i} \left( 2 \nu_i^{(k,t+1)}-2\nu_j^{(k,t+1)}  \right. \\
					&\hspace{-30pt} \left.- \nu_i^{(k,t)} + \nu_j^{(k,t)}  -  \lambda_i^{(k,t)} + \lambda_j^{(k,t)} - \nabla_{\lambda_i}\phi(\by^{(k,t)})   \right)\Big) \Big].		
				\end{split}		
			\end{align}}
		\end{enumerate}
		\textbf{-{Coordinator:}}
		\begin{enumerate}[leftmargin=*]
			\item Set $\by^{(k,t+1)} \leftarrow (\bx^{(k,t+1)}, \blambda^{(k,t+1)}, \bnu^{(k,t+1)})$.
			\item Communicate $\nabla_{\omega_i} \phi(\by^{(k,t+1)})$ to each agent $i\in\mc I$. 
			\item If the following is satisfied,
			\begin{equation}\label{eq:alg_stopping_crit}
				\|\by^{(k, t+1)} - \by^{(k,t)}\|_{\Phi} \leq (1-\beta) \varepsilon_k,
			\end{equation}
		  terminate inner iteration. Each agent then sets 
		  	\begin{equation}\label{eq:alg_update_omega}
		  \omega_i^{(k+1)} = y_i^{(k, t+1)} .
		\end{equation}
		\end{enumerate}
	\end{enumerate}
\end{algorithm}
 The linear convergence of {$\mc T^{\text{pFB}}_{{k}}$} guarantees that, for $\varepsilon_k>0$, {the iterate $\bomega^{(k+1)}$ in Proposition \ref{le:Tikhonov_alg}} is found within a finite number of {inner} iterations, {whose termination {can be} based on}  a simple stopping criterion {derived from} \eqref{eq:error_bound}. The proposed method, which results from the expansion of the pFB operator, is illustrated in Algorithm \ref{alg:FBF_tich}, {where we use $t$ as the inner iteration index.}
\begin{proposition}
	\label{prop:alg1}
	Let $(\bomega^{(k)})_{k\in\N}$ be generated by Algorithm \ref{alg:FBF_tich}. Under Assumptions  \ref{as:local_objective}--\ref{as:stepsizepFB},  for each  $k$, $\bomega^{(k+1)}$ {is an $\varepsilon_k$-approximate solution of \eqref{eq:regularized_VI} with $\bomega=\bomega^{(k)}$} and, if $\varepsilon_k>0$, the condition in \eqref{eq:alg_stopping_crit} is verified in a finite number of steps.
\end{proposition} 

\begin{remark}
	{By Proposition \ref{le:Tikhonov_alg},} Assumption \ref{as:stepsize_regul} requires {the algorithm to eventually seek an exact solution of a VI.} This is a stringent requirement, as the pFB algorithm only achieves {such {a} solution} asymptotically. The authors of \cite{facchinei_vi-constrained_2014}, alternatively, consider the {less restrictive} condition $ \lim_{k\xrightarrow{}\infty}\tfrac{\varepsilon_k}{\gamma_k} = 0$ in the case {where} the definition set of the VI-constrained VI is compact.  {Although $\bOmega$ does not satisfy this condition as the dual variables belong to an unbounded set, in practice the dual variables are bounded {in view of} \cite[Prop. 3.3]{auslender_lagrangian_2000}. }
\end{remark}

\section{{Theoretical} {connection} {and comparison} with the Hybrid Steepest Descent Method}

{In our previous works \cite{benenati_optimal_2022, benenati_optimal_2022-1}, we take a different algorithm design path,  where we reformulate Problem \eqref{eq:GNE_selection_problem} {as a} fixed point selection {problem.} This reformulation allows {one} to use}   the hybrid steepest descent method (HSDM) \cite{yamada_hybrid_2005} {to solve \eqref{eq:GNE_selection_problem}. Specifically, one {has to} find a quasi-shrinking} mapping $\mc T$ {\cite[Def. 1]{benenati_optimal_2022-1}}  such that $\fix(\mc T) = \zer(\mc T^{\text{KKT}})$. {Then,} the limit point of the sequence $(\bz^{(k)})_{k\in\N}$, defined by
\begin{equation}
	\label{eq:HSDM}
	\bz^{(k+1)} = \mc T(\bz^{(k)}) - \gamma_k \nabla\phi(\mc T \bz^{(k)}),
\end{equation}
converges to the solution of \eqref{eq:GNE_selection_problem} if $(\gamma_k)_{k\in\N}$ is square-summable but non-summable \cite[Thm. 5]{yamada_hybrid_2005}. The vanishing weight on $\nabla\phi$ is reminiscent of the Tikhonov regularization introduced in {Section \ref{sec:eq_select}.} Indeed, the two methods are related, as shown next. 

Let us consider the {exact solution to the VI in \eqref{eq:regularized_VI} with $\bomega=\bomega^{(k)}$.} Then, from {\cite[Prop. 12.3.6]{facchinei_finite-dimensional_2003}},
\begin{equation}
	\bomega^{(k+1)} = \mc J_{\frac{1}{\alpha}(\mc A + \mc B + \mc C + \gamma_k\nabla\phi)}(\bomega^{(k)})=: \mc T^{\text{Tik}}_k(\bomega^{(k)}). \label{eq:tikh_update}
\end{equation}
 The properties of the HSDM update in \eqref{eq:HSDM} depend {on} the choice of $\mc T$. Let us consider the particular case $\mc T = \mc J_{\mc A + \mc B + \mc C}$, which is averaged and, thus, {can be shown to be quasi-shrinking by \cite[Lem. 1]{benenati_optimal_2022-1}.} We rewrite \eqref{eq:HSDM} with this particular choice of $\mc T$  as:
\begin{align}\label{eq:hsdm_rewritten}
	\begin{split}
&	\bs v^{(k+1)} = \mc J_{\mc A + \mc B + \mc C}(\bz^{(k)}), \\
	&\bz^{(k+1)} = \bs v^{(k+1)} - \gamma_k \nabla\phi(\bs v^{(k+1)}),
	\end{split}
\end{align}
where we introduced the auxiliary sequence $(\bs v^{(k)})_{k\in\N}$. By rearranging \eqref{eq:hsdm_rewritten}, we note that this sequence evolves as
\begin{equation}
	\bs v^{(k+1)} = \mc J_{\mc A + \mc B + \mc C} \circ (\Id - \gamma_k\nabla\phi)(\bs v^{(k)}) =: \mc T^{\text{HSDM}}_{k} (\bs v^{(k)}). 
	\label{eq:v_sequence}
\end{equation}
	With this particular choice of $\mc T$, the operator $\mc  T^{\text{HSDM}}_{k}$ {in \eqref{eq:v_sequence}} corresponds to the forward-backward (FB) splitting method applied to the inclusion 
	$$ 0\in\mc A + \mc B + \mc C + \gamma_k\nabla\phi. $$
	From \cite[Prop. 26.1iv]{bauschke_convex_2017} and \cite[Prop. 23.38]{bauschke_convex_2017}, 
	\begin{equation*}
		\fix( \mc J_{\frac{1}{\alpha}(\mc A + \mc B + \mc C + \gamma_k\nabla\phi)}) =  \fix(\mc J_{\mc A + \mc B + \mc C} \circ  (\Id - \gamma_k\nabla\phi)),
	\end{equation*}
which implies
\begin{equation}\label{eq:same_fixed_points}
	\fix( \mc T^{\text{Tik}}_{k})=\fix(\mc T^{\text{HSDM}}_{k}).
\end{equation}
	Thus, we conclude that both the Tikhonov update in {\eqref{eq:tikh_update}} and the HSDM step in \eqref{eq:HSDM} (in terms of the auxiliary sequence $\bs v^{(k)}$) apply at each step $k$ a single update of a fixed point iteration, and the two {operators in \eqref{eq:same_fixed_points}} have the same fixed point set. {This analogy is only theoretical, as in practice the operator $\mc J_{\mc A + \mc B + \mc C}$ cannot be implemented in a distributed {fashion}.  
		Nevertheless, it outlines that both the Tikhonov method and the HSDM function by the same underlying principle of tracking the solutions to a sequence of regularized problems. Moreover, we note that \eqref{eq:same_fixed_points} does not hold for a generic choice of $\mc T$, thus the HSDM includes algorithms that are not covered by the Tikhonov method. On the other hand, clearly the operator $\mc T^{\text{Tik}}_{k}$ cannot be rewritten in terms of an FB operator. Therefore, we {should} conclude that neither method is a generalization of the other. 
		
	{The key differences between the two frameworks are summarized in Table \ref{tab:comp}. In addition, we remark that the Tikhonov framework can be paired with any (splitting) methods for {strongly monotone} games to obtain a decentralized algorithm. Meanwhile, the HSDM requires methods for monotone games that are quasi-shrinking, such as the forward-backward-forward (FBF) splitting \cite[Sec. 26.6]{bauschke_convex_2017}. Thus, Tikhonov-based methods can benefit from a larger pool of available algorithms for future development.} }
	
	{
		\begin{table}
			\centering
			\begin{tabular}{l |  c  c  c}
				& \multirow{2}{*}{Tikhonov} & Tikhonov& \multirow{2}{*}{HSDM}\\
				& &  (bounded set) &  \\
				\hline
				parameters &$\sum \gamma_k = \infty$, & $\sum \gamma_k = \infty$, & $\sum \gamma_k = \infty$,\\
				& $\varepsilon_k=0$  & $\lim_{k\to \infty}\hspace{-2pt}\frac{\varepsilon_k}{\gamma_k}\hspace{-2pt}=\hspace{-2pt}0$ & $\sum \gamma_k^2 < \infty$\\
				\hline
				$t$ (\#inner  & $\infty$ & \multirow{2}{*}{$t\to \infty$} & \multirow{2}{*}{$1$}\\
				iterations)  & eventually& & \\	
				\hline
				$\phi$ coercive & yes & no & no \\
			\end{tabular}
			\caption{{Theoretical property differences of the Tikhonov  method and HSDM for GNE selection.}}
			\label{tab:comp}
	\end{table}}

	\section{Numerical simulations}
	\label{sec:num_sim}
	
	We test the proposed algorithm on $100$  game {instances} with $10$ agents, where the pseudogradient is in the form
	\begin{equation*}
		F(\bx) = Q_F\bx + \bs c_F.
	\end{equation*}
 {The parameters} $Q_F$, {which is singular and positive semi-definite,} and $\bs c_F$ are randomly generated. We define the selection function
\begin{equation*}
	\phi(\bomega) = \|\bx\|^2_{Q_\phi} + \bs c_{\phi}^{\top} \bx + \theta (\|\blambda\|^2 + \|\bnu\|^2),
\end{equation*}
where $Q_{\phi}\succcurlyeq 0$ and $ \bs c_{\phi}$ are as well randomly generated and $\theta=10^{-3}$. For all $i$, we define the local constraint set
$ {\mc X_i =} \{\R^5: \|x_i\|_{\infty}\leq 1\}. $
	Furthermore, we {let $A_i = I$, for all $i \in \mc I$, and $b = 2\cdot\1_5$.}
	We then set $(\gamma_k)_{k\in\N}$ and $(\varepsilon_k)_{k\in\N}$ in Algorithm \ref{alg:FBF_tich} to 
	$$ \gamma_k = 10^{-3}k^{-\xi };~~ \varepsilon_k=\begin{cases}
		10^{-3}k^{-\xi\zeta}& \text{if}~10^{-3}k^{-\xi\zeta}\geq \underline\varepsilon \\
		0 &\text{if}~10^{-3}k^{-\xi \zeta}< \underline\varepsilon 
	\end{cases}$$
where $ \underline\varepsilon $ is set to the computer numerical precision. The parameter $\xi$ controls the decay of the regularization weight $\gamma_k$, while $\zeta$ controls the decay rate of $\varepsilon_k$ with respect to $\gamma_k$. We evaluate Algorithm \ref{alg:FBF_tich} in terms of the residual computed for each {outer iteration $k$ and inner iteration $t$} as}
\begin{align*}
	\mathfrak{R}(t) = \|{\by^{(k,t)}}-\mc (\Id + \mc A)^{-1}\circ (\Id - \mc B - \mc C)({\by^{(k,t)}})\|, 
\end{align*}
and in terms of the reduction of selection function $\phi$ with respect to the value obtained by the standard FBF algorithm (without optimal selection) \cite[{Alg. 2}]{franci_distributed_2020} in Figures \ref{fig:gamma}--\ref{fig:epsilon}. {we observe} that, for decreasing values of  $\xi$, the algorithm achieves a lower selection function value and a larger residual {(cf. Figure \ref{fig:gamma}).} This {trade-off between convergence to a GNE (measured by the residual) and convergence to a $\phi$-optimal point} is expected, because a too slow decay of the regularization weight $\gamma_k$ leads the algorithm to disregard the GNE seeking in order to compute the unconstrained optimal value of $\phi$. {However, if $\xi$ is set too high, the algorithm achieves a value of $\phi$  higher than the one computed without selection.} {Moreover,} an increasing value of $\alpha$ {improves} the algorithm performance {(cf. Figure \ref{fig:alpha}).} Finally, {we observe} that for increasing values of $\zeta$, the algorithm reaches a higher residual and a lower selection function value {(cf. Figure \ref{fig:epsilon}).} In our experience, {setting $\zeta$ too high might cause convergence failure} as  $\varepsilon_k$  might become 0 before $\gamma_k$ reaches a negligible value. In Figure \ref{fig:hsdm_vs_tich}  we compare Algorithm \ref{alg:FBF_tich} {with} the HSDM paired with the FBF algorithm \cite[Alg. 1]{benenati_optimal_2022-1}. The parameters for Algorithm \ref{alg:FBF_tich} are chosen among the ones that performed reasonably well in both the performance metrics considered in Figures \ref{fig:gamma}--\ref{fig:epsilon}. We find the two algorithms to have {similar} convergence speed. One might find this surprising, as the Tikhonov method is double-layered; thus, one could expect {slower} convergence  {compared}  to the single-layer HSDM. This {can be} explained by noting that the slowdown caused by the {double-layer} iterations is compensated by the linear convergence of the pFB. {In contrast,} the HSDM {uses} the FBF, {which} only achieves sublinear convergence. {Nevertheless,} as {observed in the first set of simulations,} the Tikhonov method requires more careful {parameter} tuning {than} the HSDM.}  
  
  \begin{figure}
 	\includegraphics[width=\columnwidth]{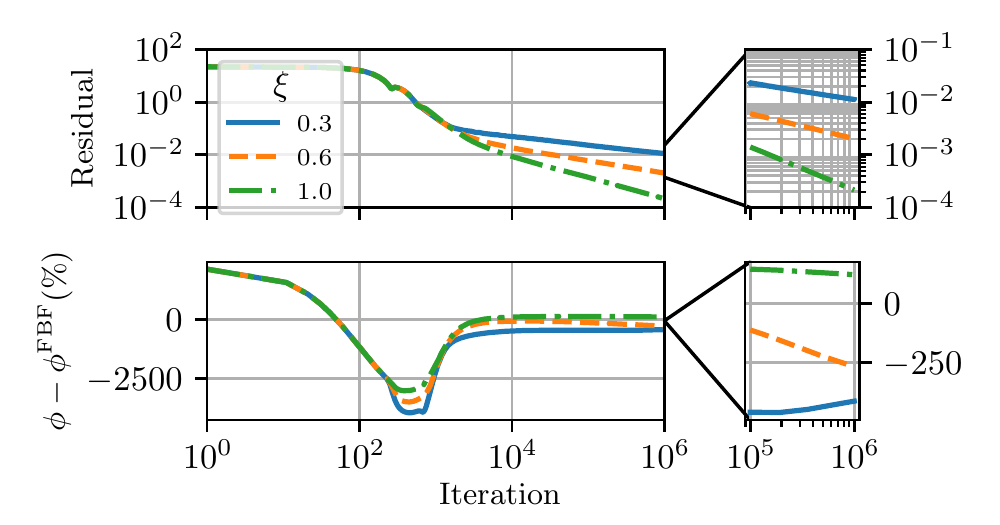}
 	\caption{{Average} performance of Algorithm \ref{alg:FBF_tich}  for $\zeta=2$ and $\alpha=1$.}
 	\label{fig:gamma}
 \end{figure}
 \begin{figure}
 	\includegraphics[width=\columnwidth]{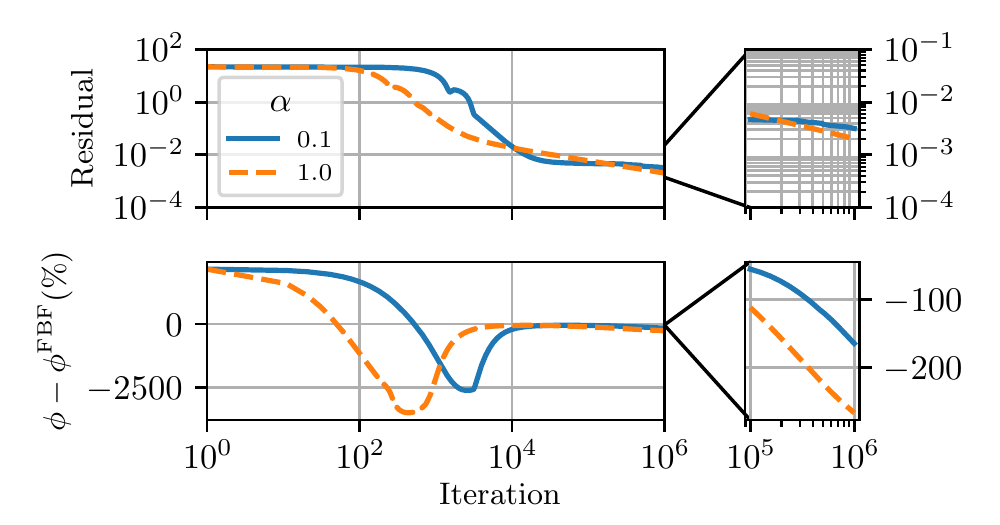}
 	\caption{{Average} performance of Algorithm \ref{alg:FBF_tich} for $\xi=0.6$  and $\zeta=2$.}
 	\label{fig:alpha}
 \end{figure}
 \begin{figure}
 	\includegraphics[width=\columnwidth]{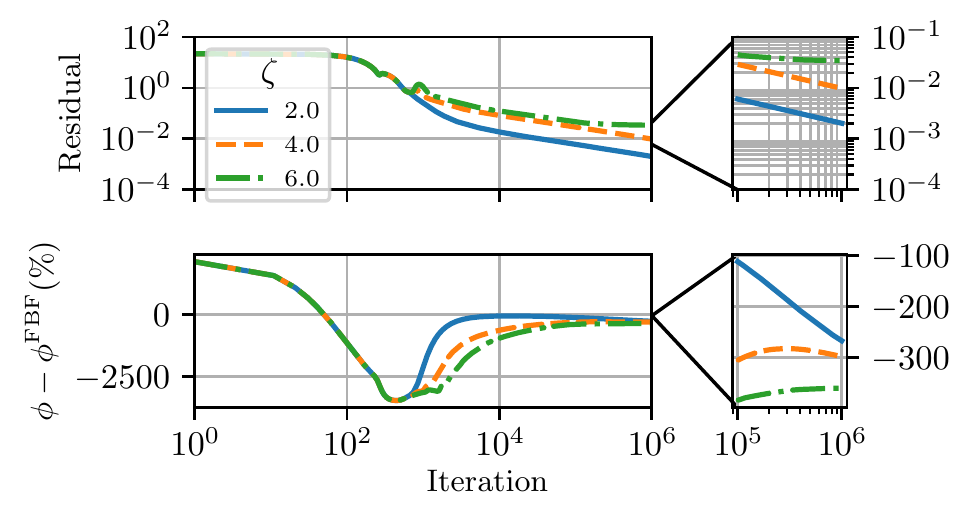}
 	\caption{{Average} performance of Algorithm \ref{alg:FBF_tich}  for $\alpha = 1$ and $\xi=0.6$.}
 	\label{fig:epsilon}
 \end{figure}
 \begin{figure}
 	\includegraphics[width=\columnwidth]{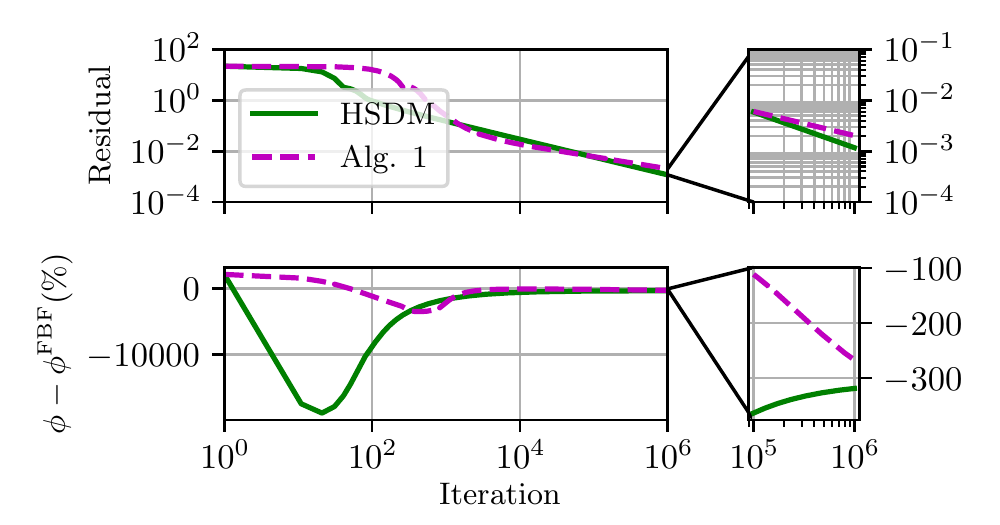}
 	\caption{Comparison between Algorithm \ref{alg:FBF_tich} with $\xi=0.6$, $\zeta=2$, $\alpha=1$ and the HSDM-FBF method \cite[Alg. 1]{benenati_optimal_2022-1}. {The x-axis indicates the cumulative number of inner iterations for Algorithm \ref{alg:FBF_tich}.} }
 	\label{fig:hsdm_vs_tich}
 \end{figure}
 
 \section{Conclusion}
 The generalized Nash equilibrium selection problem can be solved with a semi-decentralized {algorithm} based on the Tikhonov regularization method combined with the preconditioned forward backward algorithm, which achieves linear convergence for the regularized sub-problems. {Interestingly,} both the Tikhonov regularization method and a particular instance of the hybrid steepest descent method seek at each iteration an approximate solution to the same regularized problem, {{indicating} a conceptual connection.} {Although theoretically less practical ({as shown in} Table \ref{tab:comp}),} the Tikhonov algorithm {demonstrates} comparable convergence performance to the state-of-the-art {in our simulations.}

\appendix
\subsection{Proof of Lemma \ref{le:str_monotone}}
	The result 1) is immediate from the definition of strong monotonicity by applying the monotonicity of $\mc B + \mc C$ and $\nabla\phi$, which follow from \cite[Lem. 5]{yi_operator_2019} and Assumption \ref{as:phi}, respectively. From Gerschgorin's theorem, $\| \bar{\mc L}\|\leq 2\max(|\mc N_i|_{i\in\mc I})$. Thus, $\mc B$ is $\max(L_F, 2|\mc N_i|_{i\in\mc I}) $-Lipschitz continuous from Assumption \ref{as:monotonicity}. The result 2) then follows immediately.
\qedd

\subsection{Proof of Proposition \ref{le:Tikhonov_alg}}
	 {Let us denote the solution set of \eqref{eq:vi_constr_vi} by $\mc S$.}  {The proposition follows from}  \cite[Thm. 2]{facchinei_vi-constrained_2014} {if} $\bOmega$ is closed and convex; $\mc B+\mc C$ is continuous and monotone; $\nabla\phi$ is continuous and monotone plus;  {$\mc S$} is bounded and not empty; and the set 
	\begin{equation*}
		\mathfrak{L}_1	:=\{ \bomega_2 \in \bOmega | ~\exists ~\bomega_1  \in \mc S ~ \text{s. t.} ~ \nabla\phi(\bomega_2)^\top (\bomega_1 - \bomega_2)>0 \}
	\end{equation*}
	is bounded. {The {conditions on $\bOmega$ and the continuity of $\mc B, \mc C, \nabla \phi$} follow directly from the assumptions.} {The monotonicity of $\mc B + \mc C$ follows from that of skew-symmetric linear operators \cite[Ex. 20.35]{bauschke_convex_2017} as in \cite[Lemma 5]{yi_operator_2019}.} $\nabla\phi$ is monotone-plus from the convexity of $\phi$ and \cite[Ex. 22.4i]{bauschke_convex_2017}. From \cite[Thm. 2]{yi_operator_2019}, $\zer(\mc T^{\text{KKT}})\neq\varnothing$. Let $\bomega_1 \in \zer(\mc T^{\text{KKT}})$ and consider the set 
	\begin{equation*}
		\mathfrak{L}_2	:=\{ \bomega_2 \in \bOmega | \nabla\phi(\bomega_2)^\top (\bomega_1 - \bomega_2)\geq 0 \}.
	\end{equation*}
	We apply the convexity inequality on $\phi$ to find
	\begin{equation}\label{eq:convexity_ineq}
		\phi(\bomega_1) - \phi(\bomega_2) \geq  \nabla\phi(\bomega_2)^\top (\bomega_1 - \bomega_2), \quad  {\forall} \bomega_2 \in \bOmega.
	\end{equation}
	From the coercivity of $\phi$, $\phi(\bomega_1) - \phi(\bomega_2) <0 $ for  {sufficiently large} $\| \bomega_2\|$. Thus, the set $\mathfrak{L}_2$ is bounded. Therefore, from {\cite[Prop. 2.2.3]{facchinei_finite-dimensional_2003}}, $\mc S$ is non-empty and compact. {Consequently,} as \eqref{eq:convexity_ineq} holds for any $\bomega_1\in\mc S$, we find for all $\bomega_2\in\bOmega$,
	\begin{equation*}
		\max_{\bomega_1\in\mc S}  \nabla\phi(\bomega_2)^\top (\bomega_1 - \bomega_2) \leq \phi^* -  \phi(\bomega_2),
	\end{equation*}
	where  $ \phi^* = \max_{\bomega_1\in\mc S} \phi(\bomega_1)$. By the coercivity of $\phi$, $\phi^* -  \phi(\bomega_2)\hspace{-2pt}<\hspace{-2pt}0$ for {sufficiently large} $\|\bomega_2\|$. Therefore, $\mathfrak{L}_1$ is bounded.  
\qedd

\subsection{Proof of Lemma \ref{le:psi_matrix}}
	$\Phi\succcurlyeq \delta I$ follows from \cite[Lem. 6]{yi_operator_2019}. From Gerschgorin's disc theorem, 
	\begin{equation*}
		\|\Phi\| \leq  \max_{i\in\mc I}(\max(r^\mathrm{x}_i+\rho_i^{-1}, r^\lambda_i+\tau_i^{-1}, r^\nu_i+\nu_i^{-1}))\leq 2\delta.
	\end{equation*}
\qedd

\subsection{Proof of Lemma \ref{le:inner_loop_convergence}}
	The operator $(\Id + \Phi^{-1} (\mc A + \mc C))$  is non-expansive in the $\Phi$-induced norm, following \cite[Lem. 7ii]{yi_operator_2019}.  For compactness, {let us} denote $G= \mc B + \gamma_k\nabla\phi +\alpha(\Id - \bomega^{(k)})$. We proceed by proving that $ (\Id - \Phi^{-1}G)$ is contractive. First, we {observe} 
	\begin{subequations}
		\begin{align}
			&	\|\Phi\| I \succcurlyeq \Phi \succcurlyeq  \delta I \Rightarrow \|\Phi\|\|\bz\|^2 \geq \|\bz\|^2_{\Phi} \geq \delta \|\bz\|^2, \label{eq:norm_ineq_1} \\
			& \Phi \succcurlyeq  \delta I \Rightarrow  \Phi^{-1} \preceq \delta^{-1}I \Rightarrow  \|\bz\|^2_{\Phi^{-1}} \leq \delta^{-1} \|\bz\|^2, \label{eq:norm_ineq_2}
		\end{align}
	\end{subequations}
	{for any $\bz\in\bOmega$. Furthermore,}  for any pair $\bz, \bz' \in \bOmega$, 
	\begin{align}
			&\langle \bz - \bz', \Phi^{-1}(G\bz - G\bz')\rangle_{{\Phi}}\notag \\
			&\geq  \langle \bz - \bz', \Phi^{-1} \gamma_k(\nabla\phi(\bz) - \nabla\phi(\bz'))  +\alpha\Phi^{-1} (\bz - \bz')\rangle_{\Phi} \notag\\
			&=  \langle \bz - \bz',  \gamma_k(\nabla\phi(\bz) - \nabla\phi(\bz'))  +\alpha(\bz - \bz')\rangle  \notag\\
			&\geq \alpha\| \bz - \bz'\|^2 \geq \tfrac{\alpha}{\|\Phi\|} \| \bz - \bz'\|_{\Phi}^2. \label{eq:str_monotone_bound}
	\end{align}
	We {use} the fact that $\Phi^{-1}\mc B$ is cocoercive in the $\Phi$-induced norm \cite[Lem. 7i]{yi_operator_2019} (and, thus, monotone) in the first inequality,  the monotonicity of $\nabla\phi$ in Euclidean norm in the second inequality, and \eqref{eq:norm_ineq_1} in the third inequality. We then {have that}
	\begin{align} \label{eq:lipschitz_bound}
		\begin{split}
			& \| \Phi^{-1} G(\bz) -\Phi^{-1} G (\bz')  \|_{\Phi}^2 = \| G(\bz) - G (\bz')  \|_{\Phi^{-1}}^2 \overset{\eqref{eq:norm_ineq_2}}{\leq} \\
			& \tfrac{1}{\delta} \| G(\bz) - G (\bz')  \|_2^2 {\leq}  \tfrac{L_G^2}{\delta} \| \bz - \bz'   \|_2^2 \overset{\eqref{eq:norm_ineq_1}}{\leq}  \tfrac{L_G^2}{\delta^2} \| \bz - \bz'   \|_{\Phi}^2, 
		\end{split}
	\end{align}
	where we use the Lipschitz continuity of $G$ (Lemma \ref{le:str_monotone}). By expanding the square and from \eqref{eq:str_monotone_bound} and  \eqref{eq:lipschitz_bound}, {we have that}
	\begin{align*}
		&\|  (\Id - \Phi^{-1} G) (\bz)  - (\Id - \Phi^{-1} G) (\bz') \|^2_{\Phi} \\
		& = \| \bz - \bz' \|_{\Phi}^2 + \| \Phi^{-1} G(\bz) -\Phi^{-1} G (\bz')  \|_{\Phi}^2 \\
		&  \quad \ {-2} \langle \bz - \bz', \Phi^{-1}(G\bz - G\bz')\rangle_{\Phi}  \\
		&\leq  (1 + \tfrac{L_G^2}{\delta^2} - \tfrac{2\alpha}{\|\Phi\|}) \| \bz - \bz'   \|_{\Phi}^2.
	\end{align*}
	From Assumption \ref{as:stepsizepFB} and Lemma \ref{le:psi_matrix}, 
	$$  \frac{L_G^2}{\delta^2} \leq  \frac{L_G^2\alpha}{L_G^2 \delta} \leq \frac{2\alpha}{\|\Phi\|},  $$
	which implies that $\Phi^{-1}G$ is contractive in $\Phi$-norm.
	Finally, from \cite[Prop. 26.1iv]{bauschke_convex_2017} and {\cite[Eq. (1.1.3)]{facchinei_finite-dimensional_2003}}, $\fix(\mc T^{\mathrm{pFB}}_{k}) = \zer(\mc A + \mc C + G) = \SOL(\bOmega, \mc C + G)$ and the thesis follows immediately from \cite[Thm. 1.50 {(iii) and (v)}]{bauschke_convex_2017}.
\qedd

\subsection{Proof of Proposition \ref{prop:alg1}}
	For {any} $k\in\N$, denote $G_k= \mc B + \gamma_k\nabla\phi +\alpha(\Id - \bomega^{(k)})$ {and} $\by^{(k, 0)}=\bomega^{(k)}$. Let us consider the iteration 
	\begin{equation}\label{eq:algorithm_pfb_iteration}
		\by^{(k, t+1)} =\mc T^{\mathrm{pFB}}_{k} (\by^{(k, t)}),
	\end{equation}
	{and} denote $\by^{(k, t)} = (\blambda^{(k,t)}, \by^{(k,t)}, \bnu^{(k,t)} )$. From \eqref{eq:pFB}, 
	$$  (\Id - \Phi^{-1} G_k)(\by^{(k, t)})\in (\Id + \Phi^{-1}(\mc A + \mc C))(\by^{(k,t+1)}).  $$
	By multiplying each side by $\Phi$ and by substituting the definition of $\mc A, \mc C$  and $\Phi$ in the latter, we obtain
	\begin{align*}
		& (\Phi - G_k)(\by^{(k, t)})\in \Psi \by^{(k,t+1)} + \\
		& \begin{bmatrix}
			(\nc_{\X}(\bx^{(k, t+1)})) \\
			\nc_{\R^{|\mc I|m}_{\geq 0}}(\blambda^{(k,t+1)}) \\ 
			\bs 0_{|\mc I|m}
		\end{bmatrix} + \begin{bmatrix}
			\0_{{n}}\\
			b- 2\bs A \bx^{(k,t+1)} - 2\bar\L \bnu^{(k,t+1)} \\
			\0_{{|\mc I|m}}
		\end{bmatrix}.
	\end{align*} 
	By multiplying both sides by $\Psi^{-1}$, rearranging the terms and by  applying $(\Id + \nc_{C})^{-1} = \proj_C$ \cite[Ex. 23.4]{bauschke_convex_2017} for any $C$ closed, convex, it is immediate to verify that the updates in \eqref{eq:primal_update}--\eqref{eq:dual_update} are equivalent to the iteration \eqref{eq:algorithm_pfb_iteration}. From \eqref{eq:alg_stopping_crit}, \eqref{eq:alg_update_omega}, and Lemma \ref{le:inner_loop_convergence}, 
	\begin{align*}
		\|\bomega^{(k+1)} - \bomega_k^*\|_{\Phi} &= \|\by^{(k,t+1)} -   \bomega_k^*\|_{\Phi} \\
		&\leq \| \by^{(k,t+1)} -  \by^{(k,t)}\|_{\Phi}/(1-\beta)\leq \varepsilon_k.
	\end{align*}
	Thus, $\bomega^{(k+1)}$ {is an $\varepsilon_k$-approximate solution.} {Next, by triangle inequality, we have that}
	\begin{align*}
		\|\by^{(k,t+1)}- \by^{(k,t)}\|_{\Phi} &\leq \|\by^{(k,t+1)}- \bomega_k^*\|_{\Phi} +\|\bomega_k^*- \by^{(k,t)}\|_{\Phi}  \\
		&\overset{\{1\}}{\leq} (1 + \beta)\|\bomega_k^*- \by^{(k,t)}\|_{\Phi} \\
		& \overset{\{2\}}{\leq} \beta^{t} (\beta+1)\|\bomega_k^*- \by^{(k,0)}\|_{\Phi},
	\end{align*}
	where $\{1\}$ follows from the contractivity of  $\mc T^{\mathrm{pFB}}_{k}$ (see proof of Lemma \ref{le:inner_loop_convergence}) and $\{2\}$ follows from \cite[Thm. 1.50iii]{bauschke_convex_2017}. {Thus, if $\varepsilon_k>0$, \eqref{eq:alg_stopping_crit} is satisfied for 
		$$ t\geq \log_{\beta}\left(\tfrac{\varepsilon_k}{(\beta+1)\|\bomega_k^*- \by^{(k,0)}\|_{\Phi}}\right).$$} 
\qedd

	\bibliography{bibliography_bibtex}

\begin{thebibliography}{10}

\bibitem{bakhshayesh_decentralized_2021}
B.~G. Bakhshayesh and H.~Kebriaei, ``Decentralized equilibrium seeking of joint
  routing and destination planning of electric vehicles: A constrained
  aggregative game approach,'' {\em IEEE Transactions on Intelligent
  Transportation Systems}, pp.~1--10, 2021.

\bibitem{belgioioso_operationally-safe_2022}
G.~Belgioioso, W.~Ananduta, S.~Grammatico, and C.~Ocampo-Martinez,
  ``Operationally-{Safe} {Peer}-to-{Peer} {Energy} {Trading} in {Distribution}
  {Grids}: {A} {Game}-{Theoretic} {Market}-{Clearing} {Mechanism},'' {\em IEEE
  Transactions on Smart Grid}, 2022.

\bibitem{ananduta_equilibrium_2022}
W.~Ananduta and S.~Grammatico, ``Equilibrium seeking and optimal selection
  algorithms in peer-to-peer energy markets,'' {\em Games}, vol.~13, no.~5,
  2022.

\bibitem{wang_generalized_2014}
J.~Wang, M.~Peng, S.~Jin, and C.~Zhao, ``A {Generalized} {Nash} {Equilibrium}
  {Approach} for {Robust} {Cognitive} {Radio} {Networks} via {Generalized}
  {Variational} {Inequalities},'' {\em IEEE Transactions on Wireless
  Communications}, vol.~13, no.~7, pp.~3701--3714, 2014.

\bibitem{facchinei_generalized_2010}
F.~Facchinei and C.~Kanzow, ``Generalized {Nash} equilibrium problems,'' {\em
  Annals of Operations Research}, vol.~175, pp.~177--211, Mar. 2010.

\bibitem{yin_nash_2011}
H.~Yin, U.~V. Shanbhag, and P.~G. Mehta, ``Nash equilibrium problems with
  scaled congestion costs and shared constraints,'' {\em IEEE Transactions on
  Automatic Control}, vol.~56, no.~7, pp.~1702--1708, 2011.

\bibitem{kannan_distributed_2012}
A.~Kannan and U.~V. Shanbhag, ``Distributed computation of equilibria in
  monotone nash games via iterative regularization techniques,'' {\em SIAM
  Journal on Optimization}, vol.~22, no.~4, pp.~1177--1205, 2012.

\bibitem{grammatico_proximal_2018}
S.~Grammatico, ``Proximal dynamics in multiagent network games,'' {\em IEEE
  Transactions on Control of Network Systems}, vol.~5, no.~4, pp.~1707--1716,
  2018.

\bibitem{yi_operator_2019}
P.~Yi and L.~Pavel, ``An operator splitting approach for distributed
  generalized {Nash} equilibria computation,'' {\em Automatica}, vol.~102,
  pp.~111--121, 2019.

\bibitem{franci_distributed_2020}
B.~Franci, M.~Staudigl, and S.~Grammatico, ``Distributed forward-backward
  (half) forward algorithms for generalized {Nash} equilibrium seeking,'' in
  {\em Proceedings of the 2020 {European} {Control} {Conference} ({ECC})},
  pp.~1274--1279, IEEE, 2020.

\bibitem{belgioioso_distributed_2020}
G.~Belgioioso, A.~Nedić, and S.~Grammatico, ``Distributed generalized {Nash}
  equilibrium seeking in aggregative games on time-varying networks,'' {\em
  IEEE Transactions on Automatic Control}, vol.~66, no.~5, pp.~2061--2075,
  2020.

\bibitem{scutari_equilibrium_2012}
G.~Scutari, F.~Facchinei, J.-S. Pang, and L.~Lampariello, ``Equilibrium
  selection in power control games on the interference channel,'' in {\em 2012
  {Proceedings} {IEEE} {INFOCOM}}, pp.~675--683, IEEE, Mar. 2012.

\bibitem{facchinei_vi-constrained_2014}
F.~Facchinei, J.-S. Pang, G.~Scutari, and L.~Lampariello, ``{VI}-constrained
  hemivariational inequalities: distributed algorithms and power control in
  ad-hoc networks,'' {\em Mathematical Programming}, vol.~145, pp.~59--96, June
  2014.

\bibitem{benenati_optimal_2022}
E.~Benenati, W.~Ananduta, and S.~Grammatico, ``On the optimal selection of
  generalized {Nash} equilibria in linearly coupled aggregative games,'' in
  {\em 2022 {IEEE} 61st {Conference} on {Decision} and {Control} ({CDC})},
  pp.~6389--6394, IEEE, Dec. 2022.

\bibitem{benenati_optimal_2022-1}
E.~Benenati, W.~Ananduta, and S.~Grammatico, ``Optimal selection and tracking
  of generalized {Nash} equilibria in monotone games,'' Mar. 2022.
\newblock arXiv:2203.07765.

\bibitem{yamada_hybrid_2005}
I.~Yamada and N.~Ogura, ``Hybrid steepest descent method for variational
  inequality problem over the fixed point set of certain quasi-nonexpansive
  mappings,'' {\em Numerical Functional Analysis and Optimization}, vol.~25,
  no.~7-8, pp.~619--655, 2005.
\newblock Publisher: Taylor \& Francis.

\bibitem{bauschke_convex_2017}
H.~H. Bauschke and P.~L. Combettes, {\em Convex Analysis and Monotone Operator
  Theory in Hilbert Spaces}.
\newblock {CMS} Books in Mathematics, Springer international publishing, 2017.

\bibitem{palomar_convex_2009}
D.~P. Palomar and Y.~C. Eldar, {\em Convex Optimization in Signal Processing
  and Communications}.
\newblock Cambridge university press, 2009.

\bibitem{facchinei_finite-dimensional_2003}
F.~Facchinei and J.-S. Pang, {\em Finite-dimensional variational inequalities
  and complementarity problems}.
\newblock Springer series in operations research, New York: Springer, 2003.

\bibitem{auslender_lagrangian_2000}
A.~Auslender and M.~Teboulle, ``Lagrangian duality and related multiplier
  methods for variational inequality problems,'' {\em SIAM Journal on
  Optimization}, vol.~10, pp.~1097--1115, Jan. 2000.

\end{thebibliography}
	\bibliographystyle{ieeetr}

\end{document}